\documentclass[12pt]{iopart}

\usepackage{graphicx}

\begin{document}

\title{Variation of the hopping exponent in disordered silicon MOSFETs}

\author{T~Ferrus, R~George, C~H~W~Barnes, N~Lumpkin, D~J~Paul and M~Pepper}

\address {Cavendish Laboratory, University of Cambridge, J~J~Thomson avenue, CB3 0HE, Cambridge, United Kingdom}

\ead{taf25@cam.ac.uk}

\begin{abstract}

We observe a complex change in the hopping exponent value from 1/2 to 1/3 as a function of disorder strength and electron density in a sodium-doped silicon MOSFET. The disorder was varied by applying a gate voltage and thermally drifting the ions to different positions in the oxide. The same gate was then used at low temperature to modify the carrier concentration. Magnetoconductivity measurements are compatible with a change in transport mechanisms when either the disorder or the electron density is modified suggesting a possible transition from a Mott insulator to an Anderson insulator in these systems.

\end{abstract}


\submitto{\JPCM}
                        
\maketitle
  
\section{Introduction}

The crossover from Mott to Efros-Shklovskii (ES) variable-range hopping (VRH) has been extensively investigated either to confirm the existence of a Coulomb gap in 2D \cite{Efros1} or to experimentally verify its theoretical formalism \cite{Rosenbaum}. It has only been observed in the temperature dependence of conductivity of insulating devices close to the metal-to-insulator transition. Effectively, in that region, the crossover temperature, a density of state dependent entity is a slowly varying function of the electron density and disorder. This prevents a change in the value of the hopping exponent as a function of the two previous parameters from being observed. However, the existence of impurity band tails in disordered insulators \cite{Ferrus1, Ono, Hartstein} allows such a study to be performed at a fixed temperature and in the insulating regime. In this paper we show that these experimental conditions are achieved by using sodium ions in the oxide of a silicon MOSFET. While the gate controls the electron density at low temperature, the \textit{in-situ} modification of the position of the ions in the oxide by the simple application of a gate voltage at room temperature allows the disorder strength to be changed. We obtained similar results to Pepper on small silicon MOSFETs \cite{Pepper} when the ions are close to the interface and a comparable electronic situation to Ghosh's GaAs/AlGaAs devices with a fixed silicon $\delta$-doped layer \cite{Ghosh} when they are close to the metal gate. The study is extended to intermediate positions by analysing the variation of the source-drain conductivity in temperature and in magnetic field.

\section{Experiments}

All measurements were performed on identically processed silicon MOSFETs (Fig. 1a). Such devices have been widely used because of the ability to continuously vary the electron density and thus the Fermi energy by use of a metal gate. The geometry of the devices was circular (Corbino) to avoid leakage current paths around the source and drain contacts. The devices were fabricated using a high resistivity (10$^4$\,$\Omega$.cm) (100) {\itshape p}-silicon wafer to minimize the scattering with boron acceptor impurities, especially close to the silicon-oxide interface. A 40-\,nm thick gate thermal oxide was grown at 950\,$^{\circ}$C in a dry, chlorine-free oxygen atmosphere. Contacts were realized by implanting phosphorous at high dose and sputtering aluminium. Sodium ions were introduced onto the oxide surface by immersing the device in a $10^{-7}$\,N solution ($\sim 7\times 10^{11}\,$cm$^{-2}$) of high-purity sodium chloride in deionized water for 30\,s. The surface of the chip was then dried with nitrogen gas and an aluminium gate subsequently evaporated. The effective gate length of the primary Corbino MOSFET was 2\,$\mu$m and the diameter of the interior contact was 110\,$\mu$m. The Na$^+$ ions were drifted through the SiO$_2$ by applying +4\,V dc at $65^\circ$C for 10\,mins but did not diffuse into silicon\cite{SnowYon}. The device was then slowly cooled down to about 300\,mK. Once the temperature lower than about 150\,K, the ions remain \textit{frozen} in position allowing consistency in temperature dependence measurements. A series of drifts (D$i$ for the $i^{\textup{th}}$ drift) were performed and allowed modifying the position of the ions in the oxide. The last drift (D5) was carried out at $80^\circ$C for 1\,h. The undrifted device (D0) were used as a reference and for estimating the ion concentration in the oxide. Standard low-noise lock-in techniques with an amplifier gain of 10$^8$\,V/A were used to measure the source to drain conductivity. The ac excitation amplitude was $V_{\textup{ac}}$=10\,$\mu$V at a frequency of 23\,Hz. The dc offset of the amplifier was suppressed using a blocking capacitor. The gate voltage was controlled by a high resolution digital to analogue converter and the temperature measured by a calibrated germanium thermometer. All experiments were performed in an $^3$He cryostat where the magnetic field was applied perpendicular to the Si-SiO$_2$ interface.

\section{Results and discussion}

When the ions are not drifted (e.g. close to the metal gate), the variation of the source-drain conductivity with gate voltage is similar to
sodium free MOSFETs and, in the region of study (subthreshold region), the conductivity depends exponentially on gate voltage. Following a drift of the ions, the onset voltage for conduction is shifted towards negative gate voltages indicating the presence of positive charges in the oxide (Fig. 1b). Fluctuations in the conductivity also appear and strengthen as the ions approach the Si-SiO$_2$ interface. These fluctuations are reproducible both in position and height with time and with thermal cycling up to 120\,K. This result is in agreement with Pepper's results on short channel length silicon MOSFETs with low-doped substrate \cite{Pepper}. The absence of an impurity band as seen by many authors \cite{Ferrus1, Hartstein} may indicate the presence of strong potential fluctuations at the interface and a large oxide charge concentration often associated with a decrease in the impurity band visibility \cite{Timp1}. Once at the vicinity of the interface, any further drift becomes inefficient. By neglecting the ion distribution in the oxide, we can estimate the position of the ions for intermediate drifts from the experimental threshold voltage shift $\Delta V_{\textup{t}i} = V_{\textup{t}i}-V_{\textup{t}0} = e\,(d-h_i)\,N_{\textup{ox}}/\left(\epsilon_0\,\epsilon\right)$, where $N_{\textup{ox}}$ is the effective oxide charge density, $d$ the oxide thickness and $h_i$ and $V_{\textup{t}i}$ respectively the distance of the ions to the Si-SiO$_2$ interface and the threshold voltage after drift D\textit{i} (Fig. 1c). Threshold voltages were determined by the linear extrapolation to zero of $\sigma\left(V_{\textup{g}}\right)$ at 290\,mK. Allowing for this approximation, we find an upper bound for $N_{\textup{ox}} \sim 1.2\times 10^{12}\,$cm$^{-2}$ for the ion density by supposing that all the ions reached the interface after drift D5 ($h_5\sim$0). This gives a corresponding ion separation of $d_{\textup{Na}}\sim $9\,nm. The value for $N_{\textup{ox}}$ accounts for the active charges (e.g. not neutralized at the interface) and represent about 45\,\% of the total oxide charge \cite{percent}. It is an overestimate of the real ion concentration because it is plausible that some ions have already reached the Si-SiO$_2$ boundary before performing drift D5. Subsequent drifts would then change the distribution of ions at the interface by pulling the remaining ions in the oxide to the interface.

\subsection{Temperature dependence}

\begin{figure}
\centering
\resizebox{!}{6cm}{\includegraphics{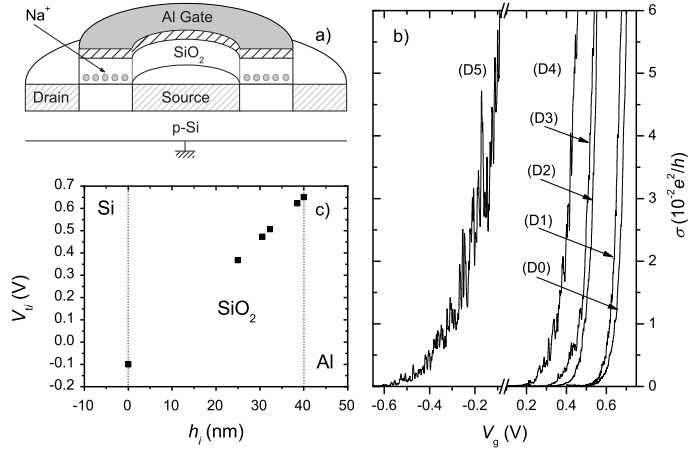}}
\caption{a) Cross-section view of a Corbino MOSFET used in the experiment after drift (D5). b) $\sigma\left(V_{\textup{g}}\right)$ for sodium ions close to the metal gate (D0) to ions close to the Si-SiO$_2$ interface (D5) at 290\,mK. c) Estimate position of the ions in the oxide for different drifts from $\Delta V_{\textup{t}i}$.}
\end{figure}

We studied the influence of the position of the ions in the oxide on the electronic properties of the device by measuring the temperature dependence of the conductivity after the different drifts. In all cases, the device showed an insulating behaviour in the region of study and the conductivity was well fitted to the generalized hopping formula over several orders of magnitude between 1\,K and 20\,K typically (Fig.\,2) \cite{note}: 
\begin{eqnarray}\label{eqn:equation1}
\sigma\left(T\right) = \sigma_0\,T^{-p s}\,\textup{e}^{-\left(T_s/T \right)^{s}}
\end{eqnarray}

\begin{figure}
\centering
\resizebox{!}{6cm}{\includegraphics{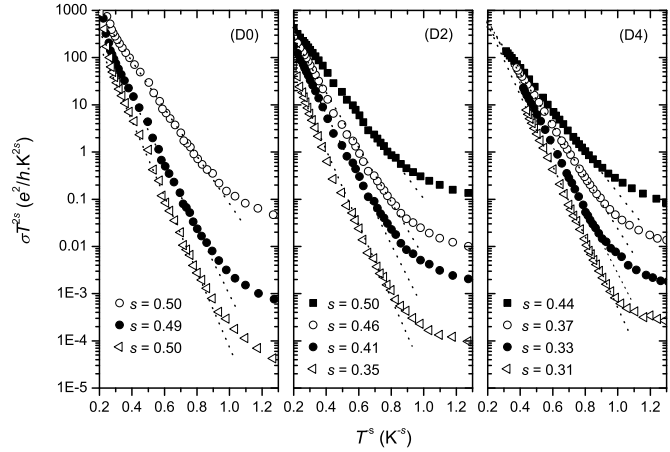}}
\caption{$\sigma\left(T\right)$ when the ions were close to the Al-SiO$_2$ interface (D0) for $V_{\textup{g}}=0.7\,$V, 0.57\,V and 0.5\,V (from top to bottom), at intermediate position in SiO$_2$ after drift D2 for $V_{\textup{g}}=0.6\,$V, 0.5\,V, 0.45\,V and 0.35\,V (from top to bottom) and close to the Si-SiO$_2$ interface after drift D4 for $V_{\textup{g}}=0.5\,$V, 0.4\,V, 0.3\,V and 0.2\,V (from top to bottom). Optimum hopping exponents are given for information. Best fits are shown with dotted lines.}
\end{figure}

\begin{figure}
\centering
\resizebox{!}{6cm}{\includegraphics{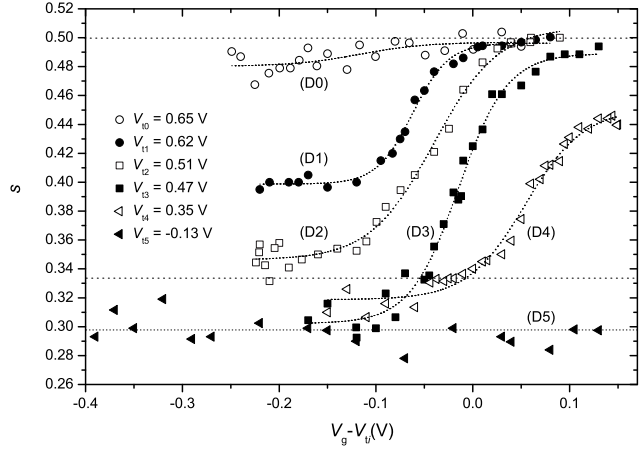}}
\caption{Variation of $s\left(V_{\textup{g}}\right)$ after after drifts D$i$.  Dotted lines represent the value of the hopping exponent in the Mott regime ($s=1/3$) and in the ES regime ($s=1/2$). The values of $s$ are determined within an error $\pm 0.002$.}
\end{figure}

Below 1\,K the conductivity is mostly determined by scattering with acoustic phonons. The fitting procedure used to find the exponent p and s is a standard one based on the minimization of the chi-square from the non-Arrhenius plot of ln$\left(\sigma T^{ps}\right)$ versus 1/$T^s$. Best fits are obtained with $p=2$ in all cases. Such a value for $p$ has already been observed in similar devices \cite{Ferrus1} as well as in Si:As \cite{Zammit} and theoretically predicted \cite{Mansfield}. When the ions are close to the metal gate, $s \sim 1/2$, indicating the presence of ES VRH \cite{Efros1} (Fig.\,3). For intermediate oxide positions, we find a smooth change from $s \sim 1/2$ (ES VRH) to $s \sim 1/3$ (Mott VRH) when decreasing the gate voltage \cite{Mott}. After the drifts D2 and D3, this crossover appears at $V_{\textup{g}}\sim 0.47$\,V and 0.45\,V respectively. This indicates the Coulomb gap progressively diminishes as the electron density is decreased. Effectively, the value of $T_{1/2}$ experimentally increases when the electron density is decreased, such that, at some point, the hopping energy in the ES regime $\sim \left(TT_{1/2}\right)^{1/2}$ becomes greater than the Coulomb gap $\sim T_{1/2}^2 / T_{1/3}$ \cite{values,Shklovskii1}. The ES VRH then changes to Mott VRH for $T > T^* \sim \xi N\left(E_{\textup{F}}\right)^2$. Because of the presence of band tails in our device, both the density of states at the Fermi energy $N\left(E_{\textup{F}}\right)$ and the localization length $\xi$ are decreasing functions of the electron density, indicating that the ES regime is more likely to be present at high electron density. This result is inconsistent with the screening of the Coulomb interaction by metallic gate as previously reported in similar devices \cite{Ferrus1, screening} and in other 2D systems where the Mott regime is expected at high electron density \cite{Khondaker}. After the drift D4, the crossover is still present although $0.32<s<0.44$. Finally, when the ions are close to the Si-SiO$_2$ interface (drift D5), hopping transport is still active but $s\left( V_{\textup{g}} \right)$ becomes non-monotonic ($0.28<s<0.32$) due to the presence of fluctuations in the conductivity. In this specific case, the conductivity was first averaged over a range $\delta V_{\textup{g}} \sim 20\,$mV that was larger than the average fluctuations in gate voltage and centred around the chosen value for $V_{\textup{g}}$. Minimization procedures were then performed. We did not observe any substantial variation of $s$ in this case but a gate voltage independent hopping exponent $s \sim 1/3$, indicating non-interacting hopping. This clearly shows that the crossover observed when varying $V_{\textup{g}}$ is progressively disappearing as the ions get closer to the Si-SiO$_2$ interface. The presence of impurities in the silicon oxide effectively creates disorder at the Si-SiO$_2$ in the form of potential fluctuations whose amplitude is proportional to $1/h_i^2$ and whose length scale varies as $h_i$ \cite{Efros2}. Close to the Si-SiO$_2$ interface, these fluctuations are short-range and are pinning strongly the electrons at the potential minima. However, this is the long-range character of the Coulomb interaction that is responsible of the formation of the soft Coulomb gap in the density of states \cite{Shklovskii2}. Thus, hopping is not-interacting and Mott law prevails in this case. The variation of $s$ with $V_{\textup{g}}$ shows a notable change, especially with the ratio $W/U$ where $W$ is the disorder strength and $U$ the on-site energy. The presence of Mott VRH at high disorder strength is compatible with $W>U$ characteristic of Anderson localization \cite{Anderson}. But, when $W<U$, a transition to a Mott insulator is expected \cite{Mott2}. So it is legitimate to investigate the possibility of an Anderson-like transition in this device \cite{Tsui}. In this case, we would expect an increase of the localization length at the transition. Still, the determination of $\xi$ from $\sigma \left(T\right)$ is difficult because of the variation in $s$ with position of the ions in the oxide and the dependence of the density of state at the Fermi level with $V_{\textup{g}}$. However, studies at constant temperature and in magnetic field allow getting access to the variation of $\xi$ independently.

\subsection{Magnetoconductivity}

In all cases, the magneto-conductivity (MC) is negative up to 10\,T, as expected for insulating materials \cite{Tokumoto,Shklovskii1} and is well described by the general equation
\begin{eqnarray}\label{eqn:equation2}
\textup{ln}\,\sigma \left(B\right) \sim -\left(B/B_0\right)^{\gamma} \mbox{ with } l_{B_0}=\left(\hbar/eB_0\right)^{1/2}
\end{eqnarray}

The use of the exponent $\gamma$ is a convenient way of assessing the non-linearity of the MC. Its value ranges from 0.6 to 2 depending on the gate voltage range and position of the ions in the oxide. It is interesting to notice that, at specific values for $V_{\textup{g}}$, the quadratic behaviour ($\gamma=2$) may extend up to 10\,T whereas at others, the linear MC ($\gamma=1$) may be present down to 0\,T (Fig. 4). The quadratic term is associated with an increase of the localization due to the shrinkage of the electron wave functions centred at the impurity site as well as interference effects due to backward electron paths \cite{Mikoshiba}. It has already been observed in similar devices where the two-dimensional impurity band was formed in the inversion layer \cite{Timp2}. It is classically associated with closed orbits and present as long as the hops between the initial and final states are enclosed in a cyclotron orbit, e.g.

\begin{eqnarray}\label{eqn:equation3}
\xi \ll {l_B}^2/r
\end{eqnarray}

where ${l_B}^2=\hbar/eB$ and $r$ is the hopping length.

\begin{figure}
\centering
\resizebox{!}{6cm}{\includegraphics{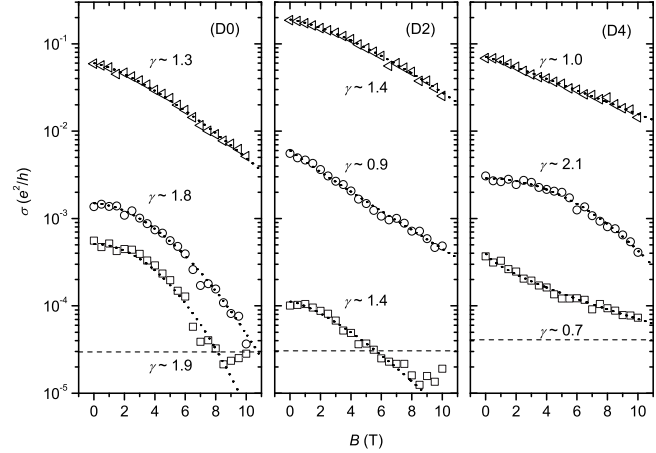}}
\caption{$\sigma\left(B\right)$ at 290\,mK for $V_{\textup{g}}$=0.7\,V, 0.58\,V and 0.55\,V (from top to bottom) after D0, $V_{\textup{g}}$=0.6\,V, 0.47\,V and 0.33\,V (from top to bottom) after D2 and $V_{\textup{g}}$=0.46\,V, 0.27\,V and 0.2\,V (from top to bottom) after D4. Best fits are shown with dotted lines. Dashed lines indicate the noise level.}
\end{figure}

\begin{figure}
\centering
\resizebox{!}{6cm}{\includegraphics{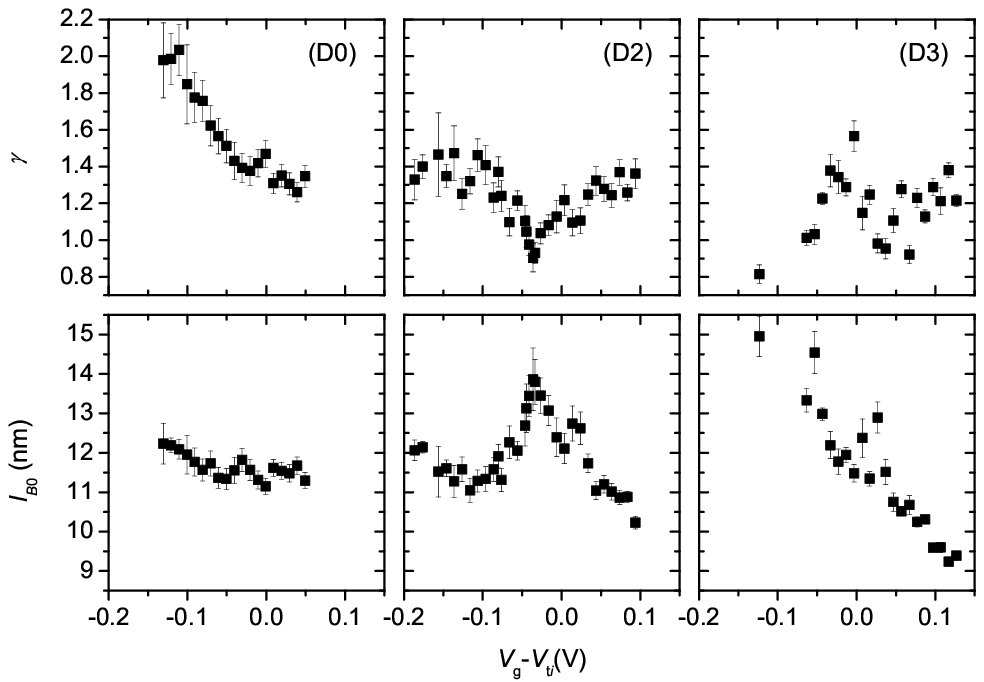}}
\caption{$\gamma\left(V_{\textup{g}}\right)$ and $l_{B_0}\left(V_{\textup{g}}\right)$ at 290\,mK for D0, D2 and D3.}
\end{figure}

This condition generally applies at low magnetic field but equivalently in a strong localized regime. As for the linear term, it has been observed experimentally by many authors \cite{Xu, Lee} and explained by the presence of strong random potential fluctuations in a two-dimensional electron gas \cite{Littlewood, Tripathi} leading to incomplete orbit scattering. In the hopping regime, this gives rise to sublinear terms in the MC under appropriate conditions \cite{Shklovskii3}. In our device both phenomena contribute to the MC, and the variation of $\gamma\left(V_{\textup{g}}, W\right)$ reflects the relative strength of the two (Fig. 5). Close to the metal-oxide interface, there is a smooth variation from a quasilinear MC for $V_{\textup{g}}>0.60\,$V to a quadratic MC for $V_{\textup{g}}<0.57\,$V. At low electron density, Eq. 3 is fulfilled and $\gamma=2$. At higher electron density, the system is no longer strongly localized and $\xi$ increases such that it invalidates Eq. 3. The distance between scattering regions diminishes giving rise to the linear or sublinear terms in the MC. When the ions are close to the silicon-oxide interface, no clear variation of $\gamma$ with $V_{\textup{g}}$ is observed. Interference effects also contribute to the fluctuations in the MC, so that even the use of averaging procedures, as described previously, cannot provide a smooth variation for $\gamma\left(V_{\textup{g}}\right)$, but surprisingly, $l_{B_0}\left(V_{\textup{g}}\right)$ is linear. At intermediate positions in the oxide, the smooth variation in $s$, as observed for D0, is not present. Instead a minimum is observed in $\gamma\left(V_{\textup{g}}\right)$ in case of drift D2 for $V_{\textup{g}} \sim 0.47\,$V, a value close to the crossover voltage in $s \left(V_{\textup{g}}\right)$. A similar but weaker minimum is also present in the case of the drift D3 at $V_{\textup{g}} \sim 0.50\,$V but none for drift D1. In presence of a magnetic field, single hops occur between impurity sites located within a cigar-shaped area of dimension $r$ and $(r \xi)^{1/2}$ \cite{Nguyen}. Because of the dependence of the hopping length on the localization length, the magnetic flux through the hopping area is proportional to $\xi^{\alpha}$ where $\alpha >0$. The magnetic field length $l_{B_0}$ as defined in Eq. 2 is thus essentially dependent on the electron density $n_{\textup{e-}}$ and the localization length \cite{Tripathi} so that a non-monotonic variation in $l_{B_0}\left(V_{\textup{g}}\right)$ could be attributed to a variation in $\xi\left(V_{\textup{g}}\right)$. In the case of drift D0, $l_{B_0}$ is nearly gate voltage independent indicating that the dependence of the hopping length on $n_{\textup{e-}}$ is weak and so $\xi\sim n_{\textup{e-}}^{3/2}$ \cite{Tripathi} whereas $\xi$ is maximum at the crossover in case of drift D2. Such an increase has been associated to the presence of an impurity band in low-doped devices \cite{Ferrus1} but in the present device, we did not observe any enhancement of the conductivity. Alternatively, it is possible to interpret an increase of $\xi$ by the occurrence of a phase transition. Because of the dependencies observed in Figs. 3 and 5, the physical process that is responsible for these is likely to rely on both the electron-electron interaction and disorder relative strengths \cite{transition}.

\section{Conclusion}

We have analysed the electronic properties of sodium-doped silicon MOSFETs by transport measurements in both temperature and magnetic field. We have observed a complex change in transport mechanism from a Mott hopping to an ES hopping regime as a function of the position of the ions in the oxide and the electron density. In gate voltage, the variation of the hopping exponent is attributed to the replenishment of the Coulomb gap due to the decrease of the localization length with electron density. This is possible because of the presence of a long impurity band tails in the density of states. This crossover only appears when the ions are within a certain distance to the silicon-oxide interface. Close to the metal gate, no crossover is observed and ES VRH is present at all sub-threshold gate voltages whereas when the ions are pinned at the silicon-oxide interface, the transport is governed by the Mott non-interacting hopping. Magnetotransport measurements showed an increase in the localization length at the crossover, suggesting the presence of a possible disorder-driven transition in these systems, like the Mott-Anderson transition.

\section*{Acknowledgement}

We would like to thank Drs F. Torregrosa and T. Bouchet from Ion Beam Services-France device processing as well as funding from the U.S. ARDA through U.S. ARO grant number DAAD19-01-1-0552. T. Ferrus is grateful to D. G. Hasko for fruitful discussions and critical reading of the manuscript.


\begin{thebibliography}{32}
\bibitem{percent} The value is estimated from the concentration of the diffused ions in the oxide using $D\,\sim 2\times 10^{-20}\,$cm$^{2}$s$^{-1}$ for the diffusion coefficient in SiO$_2$ at 300\,K
\bibitem{note} Analysis were also performed by using exponential prefactors of the form $T^{q}$ with $q$ real but fits were found to be worse in all case, especially at low electron density
\end{thebibliography}
\end{document}